\documentclass[10pt,conference]{IEEEtran}
\IEEEoverridecommandlockouts
\usepackage{cite}
\usepackage{amsmath,amssymb,amsfonts}
\usepackage{graphicx}
\usepackage{hyperref}
\usepackage{listings}
\usepackage{makecell}

\begin{document}

\title{A System for Automated Unit Test Generation Using Large Language Models and Assessment of Generated Test Suites}
\author{
\IEEEauthorblockN{Andrea Lops}
\IEEEauthorblockA{\textit{Polytechnic University of Bari \& Wideverse} \\
Bari, Italy \\
andrea.lops@poliba.it}\\
\IEEEauthorblockN{Michelantonio Trizio}
\IEEEauthorblockA{\textit{Wideverse}\\
Bari, Italy \\
michelantonio.trizio@wideverse.com}
\and
\IEEEauthorblockN{Fedelucio Narducci}
\IEEEauthorblockA{\textit{Polytechnic University of Bari}\\
Bari, Italy \\
fedelucio.narducci@poliba.it}
\\
\IEEEauthorblockN{Claudio Bartolini}
\IEEEauthorblockA{\textit{Independent Researcher}\\
claudioluisbartolini@gmail.com}
\and
\IEEEauthorblockN{Azzurra Ragone}
\IEEEauthorblockA{\textit{University of Bari}\\
Bari, Italy \\
azzurra.ragone@uniba.it}
}

\maketitle

\begin{abstract}
Unit tests represent the most basic level of testing within the software testing lifecycle and are crucial to ensuring software correctness. Designing and creating unit tests is a costly and labor-intensive process that is ripe for automation.
Recently, Large Language Models (LLMs) have been applied to various aspects of software development, including unit test generation. Although several empirical studies evaluating LLMs' capabilities in test code generation exist, they primarily focus on simple scenarios, such as the straightforward generation of unit tests for individual methods. These evaluations often involve independent and small-scale test units, providing a limited view of LLMs' performance in real-world software development scenarios.
Moreover, previous studies do not approach the problem at a suitable scale for real-life applications. Generated unit tests are often evaluated via manual integration into the original projects, a process that limits the number of tests executed and reduces overall efficiency.
To address these gaps, we have developed an approach for generating and evaluating more real-life complexity test suites. Our approach focuses on class-level test code generation and automates the entire process from test generation to test assessment.
In this work, we present \textsc{AgoneTest}: an automated system for generating test suites for Java projects and a comprehensive and principled methodology for evaluating the generated test suites. Starting from a state-of-the-art dataset (i.e., \textsc{Methods2Test}), we built a new dataset for comparing human-written tests with those generated by LLMs. Our key contributions include a scalable automated software system, a new dataset, and a detailed methodology for evaluating test quality.
\end{abstract}

\begin{IEEEkeywords}
Software Testing, Large Language Model, Automatic Assessment
\end{IEEEkeywords}

\section{Introduction}

Software testing is a critical step in the software development lifecycle, essential for ensuring code correctness and reliability. Within it, unit testing is the stage concerned with verifying the proper functioning of individual code units.
Designing and building unit tests is a costly and labor-intensive process that requires significant time and specialized skills. Automating this process represents a promising area for research and development.

Automated tools for generating unit tests can reduce test engineers' and software developers' workload. These tools typically use static code analysis methods to generate test suites. For example, EvoSuite \cite{fraser2011evosuite}, a popular tool that combines static code analysis with evolutionary search, has been demonstrated to achieve adequate coverage.

Large Language Models (LLMs), efficiently exploited in various aspects of software development, could also handle the automatic generation of unit tests.
Several empirical studies on LLMs have highlighted their ability to generate tests for simple scenarios, often limited to single methods \cite{guilherme2023initial, schafer2023empirical, siddiq2024using, yuan2023no}. Though directionally useful, these explorations focus on independent and small-scale test units, providing a limited view of LLMs' performance in real-world software development scenarios \cite{tang2024chatgpt}.
Moreover, previous studies do not approach the problem at a suitable scale for real-life examples. Generated unit tests are often evaluated via manual integration into the original projects, a process that limits the number of tests executed and reduces overall efficiency.

To address these gaps, we have developed an approach for generating and evaluating test suites that are more representative of real-life complex software projects. Our approach focuses on class-level test code generation and automates the entire process from test generation to test assessment.

In this work, we introduce \textsc{AgoneTest}, an automated system designed to generate test suites for Java projects, accompanied by a rigorous and systematic methodology to evaluate these generated test suites. Leveraging the \textsc{Methods2Test} dataset \cite{tufano2022methods2test}, we developed a new dataset specifically aimed at comparing human-written tests with those produced by LLMs.
We integrate libraries such as JaCoCo, PITest, and TsDetect to compute the metrics for test evaluation.

The main contributions of our work are as follows:
\begin{itemize}
    \item \textsc{AgoneTest}: we designed and developed a closed-loop, highly automated software system supporting the process of generation and assessment of unit tests, working at scale. This initial incarnation of the system works on real-life open-source Java projects integrating essential libraries like JaCoCo, PITest, and TsDetect;
    \item A methodology, underpinned and embodied into \textsc{AgoneTest}, for comprehensive evaluation of a variety of LLMs and relative prompting techniques and prompt schemata in the task of developing unit tests, and a set of metrics and test smells to assess the quality of the generated test suites;
    \item \textsc{Classes2Test}\footnote{\href{https://anonymous.4open.science/r/classes2test}{https://anonymous.4open.science/r/classes2test}}: An annotated open source Java project dataset extending \textsc{Methods2Test} \cite{tufano2022methods2test}, which maps focal classes to their related test classes. This extended dataset makes it possible to assess the test performance of an LLM on the entire class, rather than on a single method.
\end{itemize}

The paper is organized as follows. Section \ref{sec:background&related} sets the background and highlights differences between our work and related work. 
Section \ref{sec:overview} gives an overview of \textsc{AgoneTest} and its modules, detailing their functional scope. 
 Then, Section \ref{sec:in_practise} showcases how \textsc{AgoneTest} is applied in practice through an end-to-end example. Section \ref{sec:evaluation} addresses key research questions, presenting a first evaluation of the framework, while Section \ref{sec:Discussion} highlights insights and lessons learned from our experiments. Section \ref{sec:limitations} discusses the limitations of our approach and Section \ref{sec:conclusion} concludes the paper, outlining potential directions for future work.

\section{Background and Related Work}
\label{sec:background&related}
\subsection{Unit Test Generation}
Unit test generation is the automated process of creating test cases for individual software components, such as functions, methods, or modules. These test cases are used to independently verify the correct functioning of each unit.

Present techniques employ randomness-based \cite{csallner2004jcrasher, pacheco2007feedback}, constraint-based \cite{ma2015grt, sakti2014instance}, or search-based approaches \cite{andrews2011genetic, derakhshanfar2022basic}. The core idea behind these methods is to transform the problem into one that can be solved mathematically. For example, search-based techniques convert testing into an optimization problem, to generate unit test cases \cite{tonella2004evolutionary}. Consequently, the objective of these techniques is to generate all potential solutions and then select those that achieve better code coverage. 
EvoSuite \cite{fraser2011evosuite} works by accepting a Java class or method as input and applying search-based algorithms to generate a test suite that meets coverage criteria such as code or branch coverage. EvoSuite assesses test fitness using iterative processes of variation, selection, and optimization. Not only does it generate JUnit test cases, but it also provides a comprehensive report produced by inspecting the efficiency of the created test suite, based on metrics such as code coverage and mutation score.
One limitation of EvoSuite is that it often produces tests that lack clarity and readability \cite{grano2018empirical}. Additionally, EvoSuite can only be used on projects using Java 9 or lower, which limits its applicability to more modern Java projects (the last Java version, at the present time, is 22). Unlike EvoSuite, \textsc{AgoneTest} incorporates advanced evaluation metrics and test-smell recognition, providing a more comprehensive assessment of the quality of generated test suites and ensuring readability by leveraging human-like LLM-generated code. Moreover, \textsc{AgoneTest} supports all Java LTS versions, allowing projects built on newer versions to be tested as well, overcoming the compatibility limitations of EvoSuite. 

\subsection{Large Language Models for Test Generation}
Since the emergence of LLMs, they have been used for test suite generation. The first techniques exploiting LLMs were thought of as solutions to neural machine translation problems \cite{nie2023learning, tufano2020unit}. Such approaches work by translating from the primary method to the appropriate test prefix or test assertion while also fine-tuning the LLMs using the test generation dataset. For instance, AthenaTest \cite{tufano2020unit} optimizes BART \cite{chipman2010bart} using a test generation dataset in which the source is the primary method along with its corresponding code context, and the result is the complete test case.
AthenaTest focuses mainly on generating method-level tests by fine-tuning a single model, while \textsc{AgoneTest} shifts the focus to the generation of class-level tests. Our approach makes it possible to use up-to-date LLMs and not constrain prompt design, thereby handling more complex, real-world scenarios.
In light of the rapid evolution of instruction-tuned LLMs, the proliferation of methods for generating tests is on the rise, exploiting guided LLMs through appropriate prompts, as opposed to model fine-tuning \cite{deng2023large, xia2024fuzz4all}. Several proposals for evaluating LLMs in test suite generation have emerged. For example, \textsc{ChatTester} \cite{yuan2023no} proposes a tool for evaluating and improving LLM-generated tests based on ChatGPT.
ChatTester focuses on improving and evaluating tests generated by a specific LLM (ChatGPT), but requires human intervention to evaluate the generated code and does not provide an evaluation of class-level tests on multiple LLMs. \textsc{AgoneTest} provides support instead for a variety of LLMs and evaluates each LLM's performance on a wide range of real-life Java projects. \textsc{TestPilot} \cite{schafer2023empirical} is also focused on generating and improving tests using LLMs on JavaScript code. Although TestPilot performs an automated evaluation, it lacks wider applicability to projects other than the 25 repositories it considers in the work provided as reference here. \textsc{AgoneTest} offers far broader applicability by using a dataset of 9,410 Github repositories, and automatically integrating test libraries into them. \textsc{Cedar} \cite{nashid2023retrieval} instead proposes a prompt construction strategy based on \textit{few-shot learning} \cite{brown2020language} and the Codex model\footnote{\href{https://openai.com/index/openai-codex/}{https://openai.com/index/openai-codex/}} to generate tests. 
Cedar uses a specific prompt construction strategy, but it does not incorporate a structured mechanism to evaluate multiple LLMs and prompt techniques in a unified framework. \textsc{AgoneTest} provides this by allowing the integration and evaluation of various prompt engineering techniques and LLMs, offering a more holistic approach to test generation. Guilherme and Vincenzi \cite{guilherme2023initial} use gpt-3.5-turbo in analyzing the impact of variation in model hyperparameters. 
The study by Guilherme and Vincenzi presents an initial assessment but lacks automation in evaluating comprehensive test quality metrics like mutation coverage and test smells. \textsc{AgoneTest} goes a step further by automating these evaluations, integrating advanced metrics to provide a deeper analysis of the generated tests. Siddiq et al. \cite{siddiq2024using} offer a new proposal for evaluating tests generated using common datasets and experimenting with the use of new metrics \cite{palomba2016diffusion}. 
Although Siddiq et al. use Test Correctness (but not mutation coverage) on top of all the metrics that \textsc{AgoneTest} uses, their approach does not fully automate the test generation-execution-evaluation loop or focus on class-level tests. \textsc{AgoneTest} fills this gap by providing end-to-end automation and focusing on generating and evaluating complex, class-level test suites.

\subsection{Limits of Current Approaches in Applying LLMs to Unit Test Generation}\label{subsec:limits_current}
While promising, current approaches in applying LLMs to unit test generation exhibit several limitations:

\paragraph{Limited Scope} Current methods for evaluating how useful LLMs are in test code generation are mostly limited to generating code segments, rather than whole modules or components (e.g., whole classes in Java). Consequently, the research community lacks dedicated datasets for evaluating class-level test generation.
To the best of our knowledge,  studies often provide only punctual and anecdotal evaluations of the generated results \cite{tufano2020unit, yuan2023no, schafer2023empirical}.
\paragraph{Lack of Automation} No work has emerged in the literature that fully automates the test generation-execution-assessment loop, which is crucial for comprehensive and scalable testing \cite{deng2023large, xia2024fuzz4all, schafer2023empirical}

\paragraph{Subjective Choice of Prompts} In most cases, the choice of prompts to get LLMs to generate testing code remains subjective. There is no thorough evaluation of alternate prompting techniques compared to those initially proposed, leaving room for further exploration and optimization in prompt engineering. \cite{siddiq2024using, nashid2023retrieval, xia2024fuzz4all}.

\section{\textsc{Overview of AgoneTest}}
\label{sec:overview}
\begin{figure}
    \centering
    \includegraphics[width=\columnwidth]{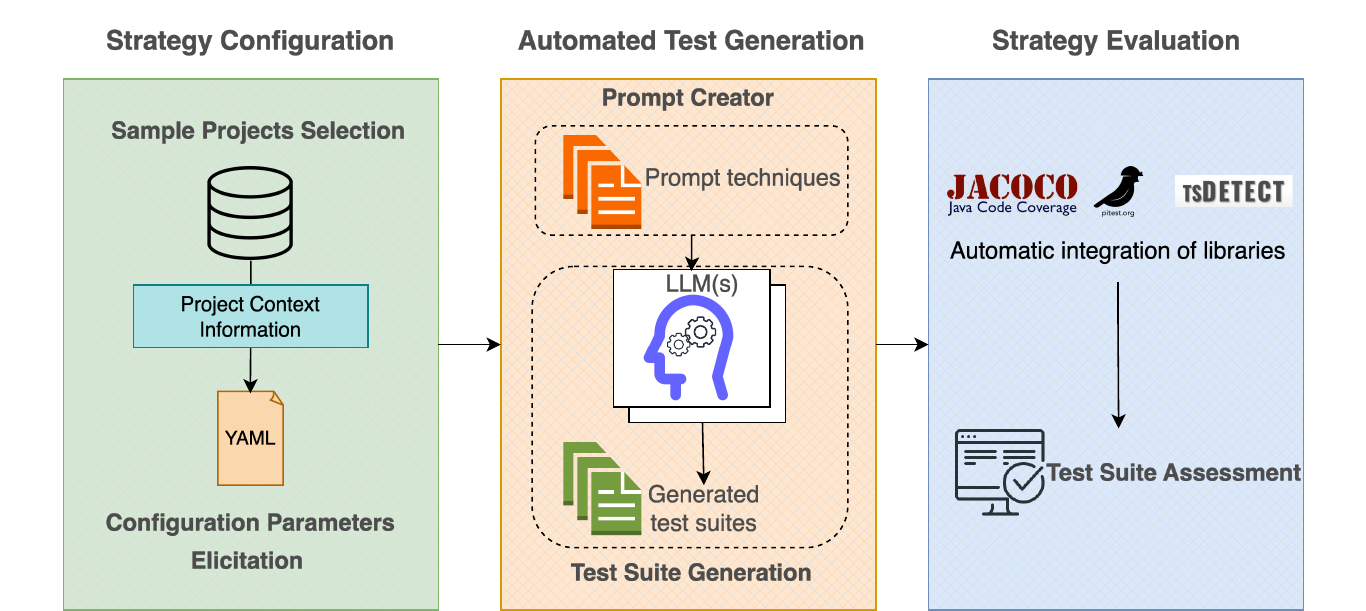}
    \caption{Overview of \textsc{AgoneTest} framework}
    \label{fig:overview}
\end{figure}

The term \textit{agone}, originating from ancient Greece and Rome, signified a contest wherein philosophers debated their ideas, with the audience determining the victor. We adopt the term \textit{agone} metaphorically to represent the competitive evaluation of LLMs and their respective prompting strategies within an arena aimed at generating optimal unit test suites. \textsc{AgoneTest} determines the optimal strategies based on standard test quality metrics, which we elaborate on in subsequent sections.

\textsc{AgoneTest} is designed to provide software testers with a system for generating and assessing unit tests. This assessment focuses on key metrics such as code coverage and the presence of known test smells, thereby offering a comprehensive assessment of test suite quality.

\textsc{AgoneTest} operates on the principle that the evaluation of LLMs in the task of generating high-quality unit tests can be performed through the collaboration of test engineers and data scientists (or prompt engineers). However, in practice, a single experienced test engineer familiar with generative AI can perform both roles, allowing the focus to be only on defining new prompt techniques and the comparison of LLMs. This is the persona that we evoke when we refer to the \textsc{AgoneTest} user (alternatively "the test engineer") in the remainder of this paper. 

The system helps test engineers through the following phases:
\begin{itemize}
    \item \textbf{Strategy Configuration}
    \item \textbf{Automated Test Generation}
    \item \textbf{Strategy Evaluation}
\end{itemize}

Figure \ref{fig:overview} provides a high-level diagram of the architecture of \textsc{AgoneTest}, showing the operating modules that streamline the test generation and evaluation process. These modules are described as follows:

\textbf{Sample Projects Selection (Strategy Configuration - I)}: As an initial configuration step, the user chooses which repositories to generate test suites for. This initial phase leverages a comprehensive dataset of annotated open-source Java repositories, which we contribute to the community. It involves preparing, loading, and managing the repositories to be tested by the system.

\textbf{Configuration Parameters Elicitation (Strategy Configuration - II)}: In this phase, configuration parameters are elicited from the selected repositories (e.g., the project java version, used testing framework, etc.) and processed to create prompts templates for the LLMs.

\textbf{Prompt Creation (Automated Test Generation - I)}: During this phase, the prompt templates used in the previous phases are fully instantiated and then used to generate unit test suites in the next step.

\textbf{Test Suite Generation (Automated Test Generation - II)}:  Here, \textsc{AgoneTest} orchestrates the interaction with the selected LLMs, feeding them the instantiated prompts to produce the unit test code. Each LLM generates test classes that are then integrated into the project structure.

\textbf{Test Suite Assessment (Strategy Evaluation)}: This phase assesses the quality of the test suites computing various metrics and identifying test smells.

This assessment enables a detailed analysis of the effectiveness and quality of the automated test generation strategies.

In the following, we describe each phase of the process in detail.

\subsection{Sample Repositories Selection}
We contribute a comprehensive, annotated dataset of open-source Java repositories from GitHub\footnote{\href{https://github.com}{https://github.com}}, which we leverage in this phase. Unlike popular datasets in the literature, our dataset enables the generation and validation of unit tests at the Java class level, rather than at the individual method level.
To collate and annotate our dataset, we built on \textsc{Methods2Test} \cite{tufano2022methods2test}. 
\textsc{Methods2Test} is a supervised dataset consisting of several Java methods, called \textit{focal methods}, univocally mapped to their respective test methods. It is a real-world dataset built using data from 9,410 open-source git repositories selected from an initial analysis of 91,385 repositories \footnote{We point out that the repositories have received updates in the last five years and are not forks, ensuring that they are actively maintained and representative of real-world software development practices.}. The dataset comprises 780,944 instances.
We chose \textsc{Methods2Test} as the starting point, as it contains not only the test methods to be tested, but also the corresponding test methods written and validated by humans. Human-written tests are a valid evaluation benchmark to evaluate the effectiveness of different LLMs in building a test suite.

We developed \textsc{Classes2Test} to enable automatic performance evaluation of LLMs in generating unit tests for entire Java classes, rather than individual methods, since this scenario is more representative of real-world applications. To accomplish this, we extracted all references to the open-source repositories present in \textsc{Methods2Test} in order to map the Java classes, referred to as \textit{focal classes}, to their corresponding test classes.

Here is the process we followed to create \textsc{Classes2Test}:
\begin{enumerate}
    \item Extract the repository reference, Github URL and selected branch;
    \item Select the classes considered in \textsc{Methods2Test};
    \item Clone the repository and save the commit hash;
    \item Map and save the focal classes along with their respective test classes.
\end{enumerate}

\begin{table}
\centering
\caption{Characteristics of the \textsc{Classes2Test} Dataset}
\begin{tabular}{|l|r|}
\hline
\thead{Characteristic} & \thead{Value} \\
\hline
Total Test Classes & 147,473 \\
Total Unique Repositories & 9,410 \\
Total Lines of Code & 173,736,517 \\
Total Cyclomatic Complexity & 81,566,509 \\
\hline
\end{tabular}
\label{tab:dataset_characteristics}
\end{table}

The resulting dataset contains 147,473 test classes extracted from 9,410 unique repositories. A summary of the dataset's characteristics is shown in Table \ref{tab:dataset_characteristics}.

\subsection{Configuration Parameters Elicitation}
\label{sub:user_input}
Before unit test generation can begin, the system extracts some parameters from the projects selected in the previous step. These parameters are then fed into the module that selects prompts and LLMs. 
To query the model under examination, various prompting techniques are available and can be chosen \cite{sahoo2024systematic}.

The configuration parameters include: 
\begin{itemize}
    \item \textbf{focal\_class}: This variable contains the Java class for which the test suite must be generated;
    \item \textbf{testing\_framework}: This variable provides the name and version of the project's testing framework (e.g., JUnit 4), directly extracted from the project during execution;
    \item \textbf{java\_version}: This variable allows you to retrieve the version of Java that the project uses.
    \item \textbf{example\_focal\_class \& example\_test\_class}: These variables contain an example focal class and the corresponding test class extracted from a reference repository, useful to provide an example to the LLM if one wants to use the few-shot prompting technique;
    \item \textbf{example\_testing\_framework \& example\_java\_version}: These variables provide the information about the example repo.
\end{itemize}
See Section \ref{subsec:LLMs-selection} for an example of a real implementation. 

\subsection{Prompt Creation}
\label{sub:prompt_creation}

In this phase, the prompt templates described in the previous phases are fully instantiated to create viable prompts to guide the LLM in generating unit tests.
We populate the user-supplied prompt structures by replacing the variables outlined in Section \ref{sub:user_input}.

It has to be noted that, in order to make sure our experiments and findings are reproducible, we prepared \textsc{Classes2Test} by saving the commit hashes of the repositories used as sources. This allows \textsc{AgoneTest} to consistently extract information such as the Java version used, the type of test framework (e.g., jUnit), and its version.

Unlike previous approaches to creating unit testing with LLMs that require human intervention to input context information \cite{yuan2023no, guilherme2023initial}, \textsc{AgoneTest} automates the process to a far greater degree. \textsc{AgoneTest} employs ElementTree \cite{garabik2005processing} and a parser to read and modify the Maven and Gradle build (see Section \ref{subsub:lib_integration}). It analyzes the libraries present and the Java version used in each build system. This method, along with the ability to use examples, offers users a versatile system for generating prompts.

\subsection{Test Suite Generation}
At this point in the process, we have everything we need for the selected LLMs to generate test suites for each focal class of the project. To ensure each model has an appropriate number of tokens, we use tiktoken\footnote{\href{https://github.com/openai/tiktoken}{https://github.com/openai/tiktoken}}, a BPE tokenizer \cite{berglund2023formalizing}, to evaluate the token count in the prompt. If the limit is exceeded, \textsc{AgoneTest} returns an error to the user, specifying the number of tokens exceeded.

We remark that \textsc{AgoneTest} allows users to evaluate a wide range of LLMs automatically. This capability is provided by the open-source LiteLLM library\footnote{\href{https://github.com/BerriAI/litellm}{https://github.com/BerriAI/litellm}}, which facilitates communication with more than 100 models\footnote{\href{https://docs.litellm.ai/docs/providers}{https://docs.litellm.ai/docs/providers}} using a standard interaction based on the OpenAI API format\footnote{\href{https://platform.openai.com/docs/guides/text-generation/chat-completions-api}{https://platform.openai.com/docs/guides/text-generation/chat-completions-api}}. Integration is made easier by LiteLLM, which translates inputs to satisfy the unique endpoint needs of each provider. This is crucial in today's environment, where the absence of standard API specifications for LLM providers makes it challenging to incorporate several LLMs into projects.

After invoking the LLM, \textsc{AgoneTest} selects relevant information from the LLM's answer (i.e., the generated test class). This step is crucial for automating the entire process, since LLMs can provide detailed descriptions or explain how the code should be structured without actually generating it. \cite{chen2024chatunitest}. In this component, \textsc{AgoneTest} removes unnecessary parts (like outline descriptions) and creates a new file to integrate the test class into the project.


\subsection{Test Suite Assessment}
\label{subsection:test-evaluation}
Here we evaluate the quality of the test suite according to the quality metrics and the test smells described below. The actual determination of metrics and test smells is done via library integration, allowing for fully automated test suite assessment.
It is important to note that this component is separate from the experimental evaluation discussed later. Instead, it serves as an additional tool provided by \textsc{AgoneTest} to assist engineers in assessing the quality of the generated tests.

\subsubsection{\textbf{Coverage Metrics}}
 \begin{itemize}
     \item \textbf{Line coverage} \cite{aniche2022effective}: This metric measures the percentage of lines of code executed during the testing process. A 100\% line coverage means that every line of code in the software has been run at least once during testing. We selected it because it provides direct visibility over the portion of the source code that is being tested.
     \item \textbf{Method coverage} \cite{aniche2022effective}: Similar to line coverage, this metric focuses on the specific methods or functions in the code. A 100\% method coverage score means that all methods have been run at least once during testing. This metric is useful to identify methods that may not have been adequately tested.
     \item \textbf{Branch coverage} \cite{aniche2022effective}: This metric calculates the percentage of decision points (such as \texttt{if} or \texttt{switch} statements) that have been executed in tests. It ensures that all possible paths in the code are tested, which can uncover defects that might be missed by line or method coverage alone.
     \item \textbf{Instruction coverage} \footnote{\href{https://www.eclemma.org/jacoco/trunk/doc/counters.html}{https://www.eclemma.org/jacoco/trunk/doc/counters.html}}: This metric calculates the number of Java bytecode instructions executed during testing. It is a detailed metric, unaffected by source code formatting, and can be determined even without debug information in the class files. This helps in pinpointing the smallest code fragments not covered by tests.
     \item \textbf{Mutation coverage} \cite{aniche2022effective}: This metric evaluates the effectiveness of tests in identifying deliberately introduced changes (mutations) in the code, such as modifying an arithmetic operation or reversing a condition. If the tests detect all mutations (i.e., identify all changes), the mutation coverage score is 100\%. This metric was chosen because it measures the robustness of the test suite.
   \end{itemize}
\subsubsection{\textbf{Test Smells\texorpdfstring{\cite{palomba2016diffusion}}{}}}
These are indicators of inefficient or problem patterns that could negatively affect the maintainability and effectiveness of the test code. Identifying test smells helps improve the quality of the test code over time and raises awareness of potential issues in test design.
\textsc{AgoneTest} determines whether the following test smells are present in the code:
 \begin{itemize}
     \item Assertion Roulette (AR) \cite{van2001refactoring}: indicate the number of test methods containing more than one assertion statement without an explanation/message (parameter in the assertion method);
     \item Conditional Test Logic (CTL) \cite{meszaros2003test}: indicate the number of test methods that contain one or more control statements (i.e., \texttt{if}, \texttt{switch}, conditional expression, \texttt{for}, \texttt{foreach} and \texttt{while} statement);
     \item Constructor Initialization (CI) \cite{peruma2019distribution}: indicate if the test class contains a constructor declaration;
     \item Default Test: indicate if the test class is named either `ExampleUnitTest' or `ExampleInstrumentedTest';
     \item Duplicate Assert (DA) \cite{peruma2019distribution}: indicate the number of test methods that contain more than one assertion statement with the same parameters;
     \item Eager Test (EA) \cite{van2001refactoring}: indicate the number of test methods containing multiple calls to multiple production methods;
     \item Empty Test (EM) \cite{peruma2019distribution}: indicate the number of test methods that do not contain a single executable statement;
     \item Exception Handling (EH) \cite{peruma2019distribution}: indicate the number of test methods that contain either a throw statement or a catch clause;
     \item General Fixture: is 1 if not all fields instantiated within the \texttt{setUp} method of a test class are utilized by all test methods in the same test class;
     \item Ignored Test (IT) \cite{peruma2019distribution}: indicate the number of tests methods that contains the \texttt{@Ignore} annotation;
     \item Lazy Test (LT) \cite{van2001refactoring}: indicate the number of test methods calling the same production method;
     \item Magic Number Test (MNT) \cite{meszaros2003test}: indicate the number of test methods that contain a numeric literal as an argument;
     \item Mystery Guest: indicate the number of test methods containing object instances of files and databases classes;
     \item Redundant Print (RP) \cite{peruma2019distribution}: indicate the number of tests methods that invokes either the \texttt{print}, \texttt{println}, \texttt{printf} or \texttt{write} method of the System class;
     \item Redundant Assertion (RA) \cite{peruma2019distribution}: indicate the number of test methods that contain an assertion statement in which the expected and actual parameters are the same;
     \item Resource Optimism (RO) \cite{peruma2019distribution}: indicate the number of tests methods utilize an instance of a File class without calling the \texttt{exists()}, \texttt{isFile()} or \texttt{notExists()} methods of the object;
     \item Sensitive Equality (SE) \cite{van2001refactoring}: indicate the number of tests methods that invokes the \texttt{toString()} method of an object;
     \item Sleepy Test: indicate the number of tests methods that invokes the \texttt{Thread.sleep()} method;
     \item Unknown Test (UT) \cite{peruma2019distribution}: indicates the number of test methods that do not contain a single assertion statement and \texttt{@Test(expected)} annotation parameter.
 \end{itemize}
 
\subsubsection{\textbf{Library integration}}
\label{subsub:lib_integration}
We utilized the following libraries to compute the metrics:
\begin{itemize}
    \item JaCoCo\footnote{\href{https://www.jacoco.org/jacoco/index.html}{https://www.jacoco.org/jacoco/index.html}}: JaCoCo is a free Java library used to measure code coverage in test suite execution. It helps developers identify which parts of their code base have been thoroughly tested and which have not, facilitating a better understanding of the test coverage within the project. We selected JaCoCo because of its widespread adoption, ease of integration with build tools, and report-generation features, which are essential for metric evaluation.
    \item PiTest \cite{mariya2016comparative}: PiTest is a mutation testing system for Java and JVM-based systems. It goes beyond traditional line and statement coverage metrics in that it offers more concrete insights into the robustness of a test suite. PiTest introduces minor changes, or \textit{mutations}, into the source code and then re-runs the tests to determine whether these changes are detected. We chose PiTest because it provides a more granular and realistic view of the actual behavior and response of the system under test compared to traditional coverage tools.
    \item \textsc{tsDetect} \cite{peruma2020tsdetect}: \textsc{tsDetect} is a library that focuses on the automatic detection of test smells in software projects. Test smells refer to patterns in test code that may indicate design or implementation issues, leading to less maintainable tests and potentially hindering code comprehension. \textsc{tsDetect} was chosen for its capability to identify these smells and provide actionable guidelines for code improvement. 
\end{itemize}

In this phase, \textsc{AgoneTest} automatically includes these libraries into the project. For each run, \textsc{AgoneTest} checks the configuration files of the supported build systems (Maven and Gradle, Section \ref{sub:prompt_creation}) to determine if the necessary libraries are already present. If they are not, it modifies the configuration to add the required dependencies.

\textsc{AgoneTest} demonstrates a high degree of automation, as illustrated by its handling of the PiTest library. Specifically, if the repo uses the JUnit 5 test framework, an additional library, \textit{"pitest-junit5-plugin"}, is required. Utilizing information extracted from the repo in the Prompt Creation module (Section \ref{sub:prompt_creation}), \textsc{AgoneTest} automatically identifies the test framework in use and adds this dependency without any human intervention.

\subsubsection{\textbf{Automate Test Suite assessment}}
After adding the necessary libraries, \textsc{AgoneTest} runs a build and test to ensure there are no compilation errors. 
The test suite assessment phase of our process presents a high degree of automation, as we describe below.

\textsc{AgoneTest} generates a report with the results of the test smells and metrics computed for the LLM-generated tests. 
To achieve this, the tool automatically retrieves detailed information from the reports produced by the libraries, compiling these data for each class within each project.

This extensive computation process enables a detailed analysis of the generated test suites. By contrasting the results, the module helps identify specific strengths and weaknesses associated with each LLM and prompt configuration. It provides insights into areas where the LLMs excel and highlight potential gaps where improvements are needed.

Furthermore, this comparison facilitates a clear understanding of the nuances in how different LLMs and prompts impact the quality of test generation. It supports the identification of optimal configurations for generating high-quality tests.
This detailed analysis is crucial for refining LLMs and enhancing their capabilities in automated test generation.

By providing such in-depth evaluations, \textsc{AgoneTest} serves as a valuable tool for researchers and developers. It aids in the continuous improvement of LLMs and contributes to advancements in the field of automated testing. Ultimately, it can ensure that the tests generated are robust and reliable, improving the effectiveness of automated testing solutions.

\section{\textsc{AgoneTest} in practice}
\label{sec:in_practise}
In this section, we will demonstrate how \textsc{AgoneTest} operates in practice by describing an end-to-end run of a practical example.

We will skip the repository selection phase in our account and move straight to the configuration phase, which concerns LLM selection and prompt specification. Then, we will exemplify how the results are presented back to the user for further analysis.

\subsection{Configuration}\label{subsec:LLMs-selection}
As described in Section \ref{sub:user_input}, AgoneTest utilizes a YAML file as input, where it is possible to specify information related to two elements: \texttt{llms} and \texttt{prompts}. 
The YAML file represented in the Listing \ref{listing:1} declares usage of ‘\texttt{gpt-4}’\footnote{\href{https://openai.com/index/gpt-4/}{https://openai.com/index/gpt-4/}} and ‘\texttt{gpt-3.5 turbo}’\footnote{\href{https://platform.openai.com/docs/models/gpt-3-5-turbo}{https://platform.openai.com/docs/models/gpt-3-5-turbo}} models, both provided by OpenAI\footnote{\href{https://openai.com/about/}{https://openai.com/about/}}.

\begin{lstlisting}[
  caption=Setup of the YAML configuration file: setting of variables for two different LLMs and two different prompts.,
  label=listing:1,
  breaklines=true,
  basicstyle=\footnotesize,
]
llms:
- model: gpt-4-1106-preview
  temperature: 0
- model: gpt-3.5-turbo
  temperature: 0
prompts:
- name: zero-shot
  value:
  - role: system
    content: You are provided with Java class. Create a test class that fully tests the proposed Java class using the project information for imports. Reply with code only, do not add other text that is not code.
  - role: user
    content: "The project uses {{testing_framework}} and Java {{java_version}} and Java class is:  \n<code>\n {{focal_class}}\n</code>"

- name: few-shot
  value:
  - role: system
    content: You are provided with an example with a Java class and its test class. You are then provided with a new Java class. Take a cue from the example and create a test class that fully tests the new proposed Java class. Reply with code only, do not add other text that is not code.
  - role: user
    content: "#Example:\nThe example Java class is:\n<code>\n {{example_java_class}} \n</code>\nThe example test class is: \n<code>\n {{example_test_class}} \n</code>.\nThe Java class you must create the test for is: \n<code>\n{{focal_class}}\n</code>"
\end{lstlisting}

Prompt specification declaration consists of two sections: \texttt{name} and \texttt{value}. \texttt{name} is an identifier for labeling the type of prompt (zero-shot, few-shots, etc.). In contrast, \texttt{value} is an array of message elements of type OpenAI \footnote{\href{https://platform.openai.com/docs/api-reference/chat/create\#chat-create-messages}{https://platform.openai.com/docs/api-reference/chat/create\#chat-create-messages}}. Each individual message includes a \texttt{role} and a \texttt{content}. 

\texttt{role} can be of the following types:
\begin{itemize}
    \item \texttt{system}: This is used to instruct the model on the behavior it should adopt.
    \item \texttt{user}: This is used to indicate the request for the generation of the test class.
\end{itemize}

In the YAML configuration file, two types of prompts are specified: zero-shot and few-shot.

\subsubsection{Zero-shot}
Zero-shot refers to presenting the model with a single instance of a request or task without any previous examples for the model to draw upon \cite{radford2019language}. This method emphasizes the model's ability to comprehend and accurately execute the given task. 

\subsubsection{Few-shot}
Unlike zero-shot prompting, few-shot prompting involves providing the model with examples demonstrating the expected inputs and outputs \cite{brown2020language}. This technique aids in contextual learning by including examples in the prompt, thereby guiding the model towards improved performance. These examples serve as a conditioning for the actual request, helping the model generate more accurate and relevant responses.

This configuration file will instruct \textsc{AgoneTest} to perform the steps described in Section \ref{sub:prompt_creation}: it will fully instantiate template variables, including focal class, test frameworks used by the repos (and versions thereof), and version of the JDK used in the repos.

\subsection{Results presentation}
After running the generation phase, \textsc{AgoneTest} generates a CSV file including, for each LLM selected and each prompting technique, the metrics computed for the focal classes as well as the results about test smells.
As a way of example, Table \ref{tab:cvs} displays an extract of this file containing as well the results for the human-written tests, as they were present in the \textsc{Classes2Test} dataset. 

By examining this file, users can gain valuable insight into the strengths and weaknesses of each LLM and the prompt combination. Plus, software testers can accurately assess the effectiveness of the LLM in creating usable and effective class-level tests. How this is done is made clear in the following section, where we describe our experimental setup for validation of \textsc{AgoneTest} and discuss some results.

\begin{table*}[ht]
\caption{CSV file extract of \textsc{AgoneTest} results. The test smells are from column 10 onwards. Please refer to Section \ref{subsection:test-evaluation} for acronyms}
\resizebox{\textwidth}{!}{
\begin{tabular}{|l|l|r|l|r|r|r|r|r|r|r|r|r|r|r|r|r|r|r|r|r|r|r|r|}
\hline
\thead{model} & \thead{prompt\\name} & \thead{Project} & \thead{Focal Class} & \thead{instruction\\coverage} & \thead{branch\\coverage} & \thead{line\\coverage}   & \thead{method\\coverage} & \thead{mutation\\coverage} & \thead{AR} & \thead{CTL} & \thead{CI} & \thead{DA} & \thead{EA} & \thead{EM} & \thead{EH} & \thead{IT} & \thead{LT} & \thead{MNT} & \thead{RP} & \thead{RA} & \thead{RO} & \thead{SE} & \thead{UT} \\ \hline
gpt-3.5-turbo & zero-shot & 8313187 & LDMLPredicateParser  & 0,72 & 0,5 & 0,67 & 0,5 & 0,5 & 6 & 1 & 0 &0&0&0&1&0& 11 &11&10 &0 &0&0 &0 \\ \hline
gpt-3.5-turbo & few-shot & 8313187 & ConfigResourceBundleParser & 0,53 & 1 & 0,67 & 0,5 & 0 & 8 & 0 & 0 &0&0&1&1&0& 10 &10&13 &0 &0&0 &0\\ \hline
gpt-3.5-turbo & few-shot & 8313187 & LDMLPredicateParser & 0,72 & 0,5 & 0,67 & 0,5 & 0,5 & 5 & 1 & 0 &0&0&1&0&0& 11 &12&1 &0 &0&0 &0\\ \hline
gpt-4 & zero-shot & 8313187 & ResourceResolutionContext & 0,85 & 0,66 & 0,83 & 1 & 0,66 & 2 & 0 & 0 &0&0&0&0&0& 9 &9& 5&0 &0&0 &0\\ \hline
gpt-4 & zero-shot & 8313187 & LDMLPredicateParser & 0,88 & 1 & 0,83 & 0,5 & 1 & 3 & 0 & 0 &0&0&0&0&0& 9 &9&1 &0 &0&0 &0\\ \hline
gpt-4 & few-shot & 8313187 & ResourceResolutionContext & 0,79 & 0,64 & 0,78 & 0,96 & 0,6 & 1 & 0 & 0 &0&0&0&0&0& 9 &9&5 &0 &0&0 &0\\ \hline
gpt-4 & few-shot & 8313187 & LDMLPredicateParser & 0,72 & 0,5 & 0,67 & 0,5 & 0,5 & 2 & 1 & 0 &0&0&0&0&0& 9 &9&1 &0 &0&0 &0\\ \hline
human & - & 8313187 & ConfigResourceBundleParser & 1 & 1 & 1 & 1 & 1 & 8& 0 & 0 &0&0&0&0&0& 15 &10&0 &0 &0&0 &0\\ \hline
human & - & 8313187 & RefreshableResources & 0,87 & 0,68 & 0,87 & 0,93 & 0,66 &1& 0 & 0 &0&0&0&0&0& 4 &4&0 &0 &0&4 &0\\ \hline
human & - & 8313187 & ResourceResolutionContext & 0,8 & 0,72 & 0,81 & 0,87 & 0,69 &6& 0 & 0 &0&0&0&0&0& 15 &11&0 &0 &0&1 &0\\ \hline
human & - & 8313187 & LDMLPredicateParser & 0,88 & 1 & 0,83 & 0,5 & 1 &5& 0 & 0 &0&0&0&0&0& 4 &4&0 &0 &0&1 &0\\ \hline
... & ... & ... & ... & ... & ... & ... & ... & ... & ... & ... & ... & ... & ... & ... & ... & ... & ... & ... & ... & ... & ... & ... & ... \\ \hline
\end{tabular}
}
\label{tab:cvs}
\end{table*}

\section{\textsc{Evaluation}}
\label{sec:evaluation}
In this experimental evaluation, we aim to address the following research questions:
\begin{itemize}
\item \textbf{RQ1:To what extent is it possible to implement an automated end-to-end process for generating test suites?} We analyze the degree of automation of the framework and the points (if any) where we need the human-in-the-loop.
\item \textbf{RQ2: Can the quality of test suites automatically generated by different LLMs and prompt strategies be effectively assessed?} 
We investigate whether the framework can provide information about the quality of the test suite in terms of efficiency and robustness and help identify strengths, weaknesses, and potential improvements.
\end{itemize}

\subsection{Dataset}
In our experiment, we randomly selected 10 repositories from our dataset \textsc{Classes2Test}. These repositories contain a total of 94 focal classes of various lengths and complexity, as shown in Table \ref{tab:example_characteristics}.
The size of the sample of randomly selected repositories is chosen to be representative enough of the variability encountered in real-world projects (usually comprising of one to a handful of co-dependent repositories), while ensuring that is tractable by our system in terms of scale.

\begin{table}
    \caption{Code Characteristics of the Repositories Selected for the Experiment}
    \centering
    \begin{tabular}{|l|r|}
        \hline
        \thead{Characteristics} & \thead{Value} \\
        \hline
        Java Repositories & 10 \\
        Total Focal Classes & 94 \\
        Total Lines of Code & 189,703 \\
        Total Cyclomatic Complexity & 85,811 \\
        \hline
    \end{tabular}
    
    \label{tab:example_characteristics}
\end{table}

\subsection{LLMs and prompts configuration}
For our experiment, we selected two LLMs from the models supported by LiteLLM\footnote{\href{https://docs.litellm.ai/docs/providers}{https://docs.litellm.ai/docs/providers}}. 
We have chosen the ‘\texttt{gpt-4}’\footnote{\href{https://openai.com/index/gpt-4/}{https://openai.com/index/gpt-4/}} and ‘\texttt{gpt-3.5 turbo}’\footnote{\href{https://platform.openai.com/docs/models/gpt-3-5-turbo}{https://platform.openai.com/docs/models/gpt-3-5-turbo}} models. The ‘\texttt{gpt-4}’ model was selected for its outstanding performance on the HumanEval benchmark\cite{luo2023wizardcoder}, while ‘\texttt{gpt-3.5 turbo}’ was chosen as an earlier generation model, allowing a meaningful comparison.

In addition to choosing LLMs, adjusting their temperature parameter is crucial. This feature allows users to control the level of randomness and creativity in the generated text. By varying the temperature, users can influence the breadth and exploration of the results. High-temperature settings, such as 1.0, introduce greater randomness, resulting in more creative and diverse outcomes. Conversely, a lower temperature setting, like 0.2, produces more focused and deterministic results, leading to predictable and cautious outcomes \cite{guilherme2023initial}. By carefully adjusting the temperature parameter, users can balance innovation and coherence in text generation, ensuring the output aligns with their specific task or application requirements.
As shown in Listing \ref{listing:1}, we set the temperature to 0 in our experiment to increase the level of coherence in text generation (and to decrease the level of randomness) and make the diverse test suite generated comparable. 
Regarding prompt types, we decided to experiment with two of the most popular techniques: zero-shot and few-shot. This allows us to show the flexibility of \textsc{AgoneTest} and further explore the use of the contextual variables seen in Section \ref{sub:user_input}. 
In our experiment, the example provided in the few-shot prompt is always the same across all LLMs to ensure uniformity. The example consists of a focal class and a test class extracted from an open sample repository\footnote{\href{https://github.com/junit-team/junit5-samples/blob/main/junit5-jupiter-starter-maven/src/test/java/com/example/project/CalculatorTests.java}{https://github.com/junit-team/junit5-samples}}.

\subsection{Data collection and analysis}
Not all test suites generated from our sample repositories were valid (i.e. tests did not pass, and in some cases, the code did not even compile).
Similarly to \cite{guilherme2023initial}, we encountered some issues: some tests failed to build due to syntax errors or incorrect or non-existent imports. To move to the assessment phase, the system automatically removes test classes with errors in order to proceed with the compilation. Once all errors are removed, the system conducts a compile-and-run test to eliminate all classes that are not \textit{green suite}, i.e. where all methods are executed without failures.

This is done for two reasons: first, because after performing root cause analysis, we found that the failing test classes did so because they were incorrectly specified, rather than for the presence of bugs. For example, tests failed because they invoked private or outdated methods of libraries. Or they failed because they made calls to non-existent APIs.
The second reason is that having a \textit{green suite} is a necessary condition for PiTest to calculate mutation coverage \footnote{\href{https://pitest.org/faq/}{https://pitest.org/faq/}}. After completing this clean-up phase, the system performs a final compilation with the execution of the tests and the library to collect data for evaluation.

\begin{table}[ht]
\caption{Number of classes per experiment that compile and for which all tests are passed}
\begin{tabular}{|l|l|r|r|r|}
\hline
\thead{model} & \thead{prompt\\name} & \thead{Build} & \thead{Pass} & \thead{Total\\Rejected}\\ \hline
gpt-3.5-turbo & zero-shot & 64 (68.08\%) & 36 (38.29\%)  & 58 (61.70\%)\\ \hline
gpt-3.5-turbo & few-shot & 65 (69.14\%) & 35 (37.23\%)& 59 (62.76\%)\\ \hline
gpt-4 & zero-shot & 76 (80.85\%) & 29 (30.85\%) & 65 (69.14\%)\\ \hline
gpt-4 & few-shot& 76 (80.85\%) & 28 (29.78\%) & 66 (70.21\%)\\ \hline
human & - & 94 (100\%) & 94 (100\%) & 94 (100\%) \\ \hline
\end{tabular}
\label{tab:compilation}
\end{table}

Table \ref{tab:compilation} shows the percentage of classes that compile and for which all tests are passed.
On average, we have seen in our experiment that:
\begin{itemize}
    \item 75\% of the generated test classes compile successfully;
    \item 34\% of generated test classes are \textit{green suite} and have calculable mutation coverage.
\end{itemize}

\begin{table*}
\caption{Metrics computed for each model and prompt techniques used and for human-written tests. In bold, the best results for each metric.}
\centering
\begin{tabular}{|l|l|r|r|r|r|r|}
\hline
\thead{model} & \thead{prompt name} & \thead{instruction coverage} & \thead{branch coverage} & \thead{line coverage} & \thead{method coverage} & \thead{mutation coverage} \\ \hline
gpt-3.5-turbo & zero-shot & 0,756923077 & 0,706363636 & 0,776923077 & 0,848217345 & 0,546923077 \\ \hline
gpt-3.5-turbo & few-shot & 0,813333333 & 0,681111111 & 0,775555556 & 0,832222222 & 0,455555556\\ \hline
gpt-4 & zero-shot & \textbf{0,879090909} & 0,776923077 & \textbf{0,866363636} & \textbf{0,854545455} & 0,546363636\\ \hline
gpt-4 & few-shot & 0,753333333 & 0,775555556 & 0,781666667 & 0,836923077 & 0,461666667 \\ \hline
human & - & 0,783611111 & \textbf{0,808947368} & 0,766388889 & 0,698055556 & \textbf{0,690555556} \\ \hline
\end{tabular}
\label{tab:result}
\end{table*}

Table \ref{tab:result} shows a comparative analysis of the performance of the different combinations of LLMs and prompt techniques with respect to the metrics computed by \textsc{AgoneTest}. We also report the metrics computed for the human-written tests, to have a benchmark.  

\section{Discussion and lessons learned}
In this section, we will answer the research questions previously defined and discuss the lesson learned together with possible future research directions.

\label{sec:Discussion}
\textbf{RQ1: To what extent is it possible to implement an automated end-to-end process for generating test suites?}
\textsc{AgoneTest} provides an end-to-end automated process to generate and evaluate test suites without human intervention.
However, there are two points requiring attention that our experiments underline:
\begin{itemize}
    \item The compilation success rate of the generated test classes shows room for improvement (in our experiment ranged between 64\% and 76\%);
    \item The percentage of tests passed was relatively low (between 30\% and 38\%).
\end{itemize}

Further analysis revealed that many generated tests failed due to incorrect imports or syntax errors and not because they discovered previously undetected bugs.
To improve on these results and increase such percentages, there are different paths to explore.
One is human-in-the-loop: where human intervention might include manually fixing code errors, adjusting settings, or installing required libraries for successful execution. 

On the other hand, a good result, in light of automation, is having automated the process of extracting contextual information from project configuration files (such as Maven or Gradle). This minimizes the need for manual intervention and enhances the accuracy of the generated prompts and tests.



\textbf{RQ2: Can the quality of test suites automatically generated by different LLMs and prompt strategies be effectively assessed?} 
\textsc{AgoneTest} gives relevant information about the quality of the test suite generated, in terms of code coverage, robustness of the test suite to artificially injected bugs (i.e., mutation coverage), and test smells.
Indeed, the presence of test smells indicates potential issues with test design and maintainability.

In Table \ref{tab:result} we can see that LLMs-written test suites already have good quality in terms of coverage, but should improve in terms of robustness. 
The benchmark of human-written tests shows always better results for mutation coverage.  
Comparing the output of our LLMs-powered system against tests written by actual test engineers gave us valuable insight into each model-setup pair's capabilities and limitations.
Our experiment (Table \ref{tab:result}) demonstrates that the performance of various LLMs varied significantly depending on the prompting technique used. 
Surprisingly, we found better results for zero-shot prompts for ‘\texttt{gpt-4}’ than for the few-shot one.
However, it is important to note that this does not mean that ‘\texttt{gpt-4}’ is the best model for all scenarios, and this is neither the objective of our experiment (i.e., to find the best model). Since the performance of a model can vary significantly depending on the specifics of the context or the structure of the prompts. 
For this reason, we created \textsc{AgoneTest}: to provide users with a system that lets them experiment with various combinations to find the LLM and prompt configuration that best fits their specific requirements.

\subsection{Lessons Learned}
\label{sub:lessons-learned}

Throughout the development and evaluation of \textsc{AgoneTest}, we gathered several key insights that will guide future improvements in the framework. These lessons are crucial for refining the system and improving the efficacy of LLM-generated tests. Each subsection below highlights a specific challenge encountered and outlines a potential solution.


\subsubsection{Compilation and Test Pass Rate}
Our experiments show compilation and test pass rates that could be improved in light of the pursuit of full automation. The causes of these are diverse (e.g., import classes that do not exist or are missing). Automating the correction of these recurring problems is possible \cite{kang2023large} and will increase the success rate of the generated tests. 
One promising approach involves asking the LLM itself to analyze errors in the generated test code and provide fixes. By supplying the identified errors as feedback, the LLM can generate corrected and functional code, thereby enhancing the initial output. Additionally, enhancing the robustness of the generated tests by incorporating context-aware validation and fixing mechanisms will ensure that the test suites align closely with the project's specific structures. This integrated approach not only automates error correction but also enhances the overall reliability and effectiveness of the test generation process, moving closer to the goal of fully automated, high-quality test suite production.
 
\subsubsection{Performance in Mutation Testing}
Human-written tests consistently outperformed LLM-generated tests in terms of mutation coverage, indicating that manually written tests are more effective at identifying code changes introduced through mutation. To address this, we should focus on improving the robustness of the generated test suites by refining the prompting algorithms and incorporating mutation-aware test generation techniques.

\subsubsection{Scalability and Resource Management}
Automating the entire pipeline-downloading projects, generating test suites, integrating libraries, and performing evaluations proved to be resource intensive. Efficiently managing and parallelizing these tasks can alleviate computational overhead and improve scalability, allowing \textsc{AgoneTest} to handle larger datasets and codebases more effectively.

\subsubsection{Impact of Prompting Techniques}
The choice of prompting technique significantly impacts the quality of the generated tests. Our experiments showed that zero-shot prompting with gpt-4 yielded the best results, but performance varied across different combinations of LLMs and prompts. Systematically exploring and evaluating different prompting strategies will help identify the most effective configurations for various scenarios.

\subsubsection{Automated Context Extraction}
Providing the LLMs with accurate context information, such as the testing framework and the Java version, is essential for generating correct test classes. Automating the extraction of this context information reduces the need for manual intervention and improves the quality of generated prompts and tests. Enhancing the automation of context extraction by developing more sophisticated parsers and context inference algorithms will dynamically adapt to various project configurations.

\subsubsection{Real-world Applicability}
Building the dataset from actual open-source Java repositories on GitHub ensured that \textsc{AgoneTest} operates in real-world scenarios. However, ensuring that the dataset is representative of real-life situations across different types of repos and codebases remains an ongoing goal. To maintain and improve real-world applicability, we should continuously upgrade and update our dataset to include a broader range of real-world repositories and project structures, ensuring that the evaluation remains relevant and comprehensive.

These lessons direct us towards further improvements in \textsc{AgoneTest}. By implementing these improvements, we aim to develop a more robust, efficient, and reliable framework for automated unit test generation.

\section{Limitations}
\label{sec:limitations}
Although \textsc{AgoneTest} presents an innovative framework for automating the generation and evaluation of unit test suites using LLMs, several limitations should be acknowledged regarding its current implementation and the first experimental results.

\subsection{\textbf{Dataset and Generalization}}
For our evaluation, we relied on the newly created \textsc{Classes2Test} dataset, derived from \textsc{Methods2Test}. Although this dataset is designed to evaluate class-level test generation, its scope is limited to Java projects. This makes our findings hardly generalizable to different programming languages. Moreover, the repositories included in \textsc{Classes2Test} were selected based on their ability to compile without errors, potentially introducing a bias towards well-structured codebases.
        
\subsection{\textbf{Model and Prompt Variability}}
   \textit{Limited Number of LLMs and Prompts Tested}: Although \textsc{AgoneTest} supports various LLMs and prompting techniques, our initial experimental setup involved only two models (gpt-4 and gpt-3.5 turbo) and two prompt types (zero-shot and few-shot). As LLMs and prompt engineering techniques continue to evolve, the results might vary significantly with newer models and advanced prompts. The limited scope of our initial tests could thus restrict the breadth of our conclusions.
  
  \textit{Temperature Setting Restrictions}: In our experiment, we set the temperature parameter to 0 to ensure consistency and reproducibility. Although this reduces randomness and increases coherence, it may inadvertently limit the creativity and diversity of the generated test cases. Different temperature settings could yield more varied results, which were not explored in this study.

\subsection{\textbf{Compilation and Execution Failures}} A notable limitation observed during our experiments was the non-compilation and execution failures of some generated test classes. Approximately 66\% of the test classes generated were either rejected during the compilation phase or failed to contribute positively to the metrics due to inherent errors. This reveals the current inability of some LLMs to consistently generate syntactically and semantically correct test code, affecting the overall evaluation.
        
\subsection{\textbf{Evaluation Metrics}} While we employed a comprehensive set of metrics and test smell indicators, these metrics alone may not fully capture the quality of the test suite.\\
 


\section{Conclusion and Future Work}
\label{sec:conclusion}

In this paper, we present \textsc{AgoneTest}, a comprehensive framework for automating the generation and assessment of unit test suites using LLMs. This framework focuses on generating complex, class-level test suites while automating the entire testing process from test generation to integration and evaluation.

The results of our experiments demonstrate that \textsc{AgoneTest} can produce and evaluate unit tests across various real-world projects, offering detailed insights into the performance of different LLMs and prompting techniques. While the initial findings are promising, they also highlight challenges emphasizing the need for further refinement.

The automation of unit test generation using LLMs is a promising field. While current capabilities do not yet match those of human engineers for some tasks (such as mutation coverage), promising results in instruction, line, and method coverage indicate that further research and refinement can bridge this gap. Future work should focus on systematic research into the most effective LLMs and prompts, coupled with continuous improvements in automated correction mechanisms for recurrent issues.

\bibliographystyle{IEEEtran}
\bibliography{ms}
\end{document}